\documentstyle[11pt,newpasp,twoside]{article}
\def\edcomment#1{\iffalse\marginpar{\raggedright\sl#1\/}\else\relax\fi}
\reversemarginpar

\begin{document}
\title{Evolution of the accretion disk in the nucleus of NGC\,1097}
\author{Thaisa Storchi-Bergmann \& Rodrigo Nemmen da Silva}
\affil{Instituto de F\'\i sica, UFRGS, Porto Alegre, Brasil}

\author{Michael Eracleous}
\affil{Pensylvania State University,
University Park, PA 16802}

\begin{abstract}

We discuss the long-term evolution of the broad double-peaked
H$\alpha$ profile of the LINER/Seyfert 1 nucleus of NGC\,1097.
Besides the previously known variation
of the relative intensities of the blue and red peaks,
the profile has recently shown an increasing separation
between the two peaks, at the same time as the integrated flux
has  decreased. We successfully model
these variations using a precessing asymmetric accretion disk
with a varying emissivity law. We interpret the emissivity
variation as due to the fact that the source
of ionization is getting dimmer,
causing the region of maximum emission to drift inwards
(and thus to regions of higher velocities).
In addition, in the last 3 yrs of observations,
the central wavelength of the double-peaked line has shifted
to bluer wavelengths, which may be due to a wind
from the disk. It is the first time that such evolution is observed
so clearly, giving additional support for an accretion
disk as the origin of the double-peaked profile in NGC\,1097.

\end{abstract}

\section{Introduction}

\begin{figure*}
\vspace{7.0cm}
\caption{Spectra at four different epochs together
with best elliptical disk models.
Notice change of flux scale for the
two last epochs.}
\includegraphics{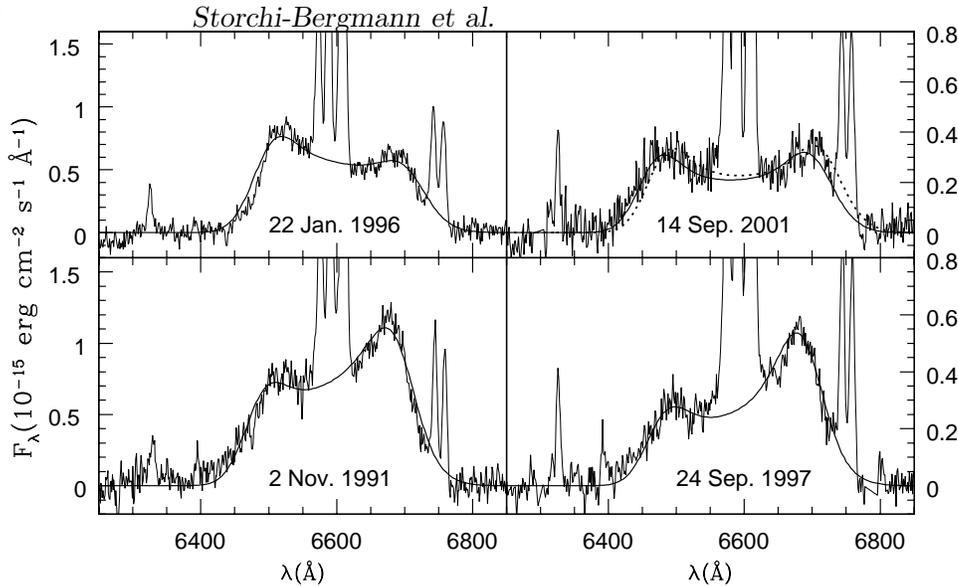}
\end{figure*}

The recent developments in the study of the stellar kinematics around the
nuclei of nearby galaxies (e.g. Ferrarese \& Merrit 2000) and in our own
support the presence of supermassive blackholes
(hereafter SMBH) in the bulges of
most present day galaxies.
And if the SMBH are there, they will eventually accrete mass, e.g. via
capture of a star passing close enough to the SMBH.
If the SMBH has a mass smaller than
10$^8$ M$_\odot$, a solar-type star will be tidally
disrupted, leading to the
formation of an accretion disk or ring (Rees, 1988).

We believe that such ring could be the source of the
broad ($\sim$10,000\,km\,s$^{-1}$) double-peaked H$\alpha$
emission which we first observed in a spectrum of the
nucleus of NGC\,1097 obtained in 1991 (Storchi-Bergmann et al. 1993).
Previous observations of the nucleus of this galaxy had only
shown narrow emission-lines typical of a LINER.
We have since then followed the variation of the
double-peaked profile and discuss, in this work,
the results of 10 yrs of monitoring.
A more detailed study of these
observations, including modeling of the ring (or disk)
will be presented elsewhere.

\section{Observed variations, interpretation and modeling}

We have observed the nucleus of NGC\,1097 using
mostly long-slit spectroscopy at the CTIO Blanco telescope,
approximately once a year.
We show in Fig. 1 the broad double-peaked H$\alpha$
profile in 4 epochs, which illustrate the main variations
observed in the profiles.

In order to quantify the variations, we
have measured the wavelengths of the blue and red
peaks($\lambda_B$ and $\lambda_R$), their peak fluxes
(F$_B$ and F$_R$) and the total flux of the broad line (F$_{broad}$).
Fig. 2 shows the variations of these quantities
as a function of time: the
wavelength of the blue peak decreases while that
of the red peak increases, the flux ratio between
the blue and red peaks varies between 0.5 to 1.5 and the
total flux of the line decreases.

\begin{figure*}
\vspace{8.0cm}
\caption{Variation of the peak wavelengths of the blue ($\lambda_B$) and red
($\lambda_R$) peaks, their flux ratio F$_B$/F$_R$ and total flux F$_{broad}$.}
\includegraphics{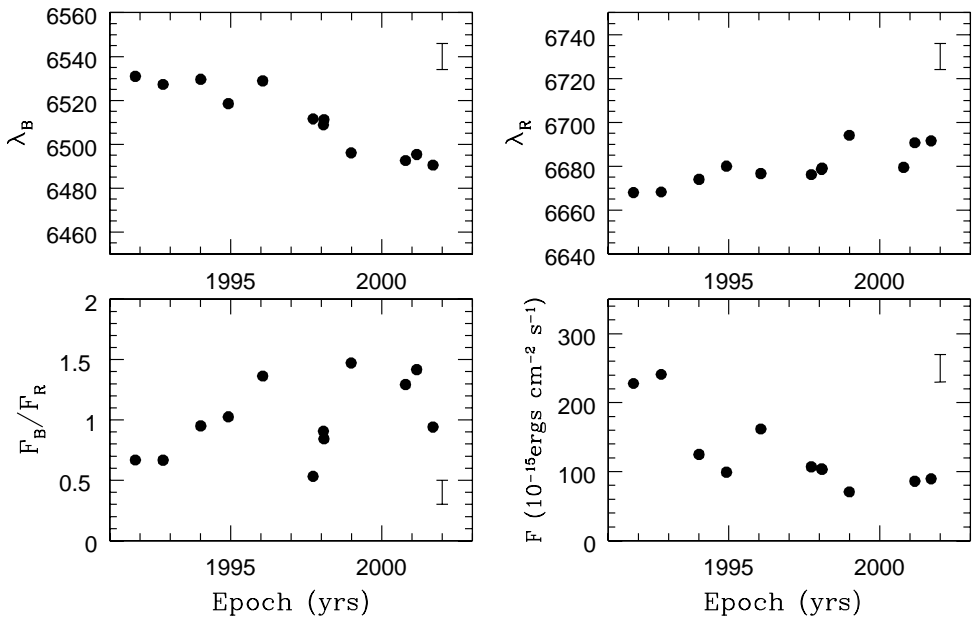}
\end{figure*}


The variation of F$_B$/F$_R$ supports the origin
of the profile on an asymmetric disk or ring, which could be the
elliptical accretion disk proposed by Eracleous et al. (1995)
or a circular disk with a disturbance like a spot or spiral arm (Gilbert et
al. 1999). The precession of the elliptical disk or of the disturbance
in the circular disk would be the origin of the variation of F$_B$/F$_R$.

\subsection{Are we witnessing the matter falling into the SMBH?}

We have used the elliptical ring model described in
Storchi-Bergmann et al. (1995, 1997)  to try to reproduce the
variations observed from 1991 to 2001, by changing only
the orientation of the major axis of the ring
relative to the line of sight $\Phi_0$. Nevertheless,
the observed shifts of the blue and red peaks could
only be reproduced by decreasing the inner radius of the ring
from 1300 gravitational radii (hereafter R$_g$) to
450 R$_g$. This result could suggest that this radius has
decreased during the ten years of observations,
and that {\it we are witnessing the matter from the disk falling towards
the SMBH!} However, such a decrease  should
occur in the viscous time scale, which is about 10$^3$ yrs (Frank
et al. 1992) in the case of NGC\,1097,
much longer than the 10 yrs spanned by our observations and the
answer to the above question is unfortunatelly {\it no}.

An alternative possibility is that the ring has
always had a smaller inner radius of 450\,R$_g$ (and
can now most properly be called a disk), but the
emissivity is changing, such that in more recent epochs
it is favoring the inner regions of the disk relative to the outer ones.
This picture is also in agreement with the observational result that
the flux of the broad line is decreasing, as shown in Fig. 2.
The emissivity law which best reproduces
the observations is a double power-law as a function of the radius,
such that the emissivity {\it increases}
from the inner radius until a radius
$\xi_q$ and then {\it decreases} beyond  $\xi_q$.
The increasing separation between
the blue and red peaks in an elliptical disk with the above
emissivity law is due to the fact that the region of maximum
emission is getting closer to the source.

The fit of this model to the profiles is shown together
with the data in Fig.\,1. These fits have evidenced another
change in the profiles beginning by 1997: we had to introduce
a blueshift of the lines of $\sim -500$\,km\, s$^{-1}$
in order to reproduce the most recent profiles. This is illustrated in
Fig. 2, which shows the fit to the Sep. 2001 profile
before (dotted lines) and after the blueshift.


\begin{figure*}
\vspace{7.0cm}
\caption{Cartoon showing differences in emissivity and orientation of
the elliptical disk between two epochs. Observer is to the right.}
\includegraphics{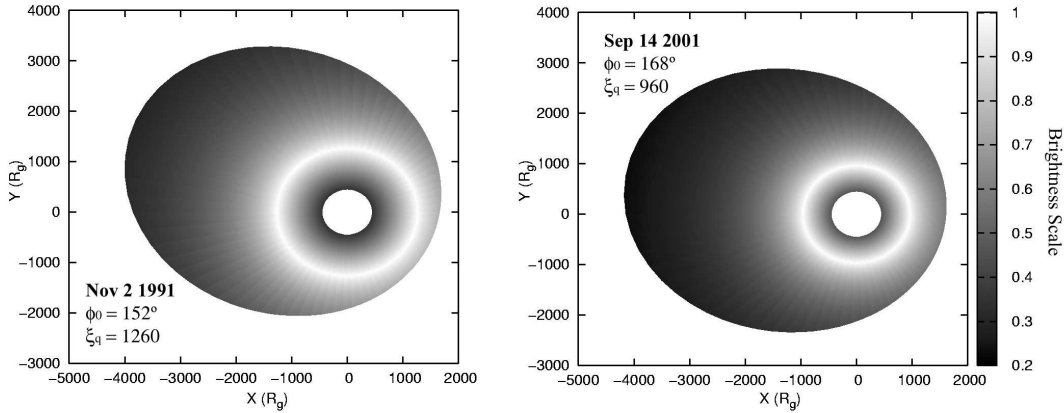}
\end{figure*}

\section{Conclusions}

The variation of F$_B$/F$_R$ supports the origin
of the profile on an asymmetric disk.
The increasing separation of the blue and red peaks indicates
that the broad-line
emission is coming from regions of successively larger velocities
and thus smaller radii,
at the same time as the broad line flux is decreasing.
The favored interpretation is that the
ionizing source is getting dimmer, causing the region of maximum emission
to drift inwards
(Dumont \& Collin-Souffrin 1990).
We have succesfuly modelled this evolution using an elliptical
disk model, with a varying emissivity law, as illustrated in Fig. 3.
The blueshift of the line which has appeared by 1997,
suggests that we are witnessing the formation of a wind,
with velocity consistent with those observed for
Fe\,II absorption lines in an HST UV spectrum
of NGC\,1097 (Eracleous 2002) and with absorption
systems in general observed in the spectra of active galaxies.

Our observations and consistent
modeling show that the double-peaked emission in NGC\,1097 present
many ingredients predicted by accretion disk models (Elvis 2000), giving
stronger support yet for its presence around a nuclear SMBH
in this galaxy.

\end{document}